\documentclass[showpacs,nofootinbib,preprintnumbers,prd,superscriptaddress,twocolumn]{revtex4}

\usepackage{tikz}

\usepackage{amsmath}
\usepackage{braket}
\usepackage{amsfonts}
\usepackage{amssymb}
\usepackage{graphicx}
\usepackage[titletoc]{appendix}
\usepackage{color}
\usepackage{hyperref}
\usepackage{cleveref}
\usepackage[rightcaption]{sidecap}
\usepackage{subcaption}

\usepackage{float}
\usepackage{wasysym}
\usepackage{tensor}
\usepackage{csquotes}

\usepackage{dcolumn}

\usepackage{array}
\usepackage{ctable}
\usepackage{multirow}
\usepackage{siunitx}
\usepackage{tabularx}
\usepackage{booktabs}

\usepackage{mathtools}
\usepackage{dirtytalk}
\DeclareSIUnit{\parsec}{pc}

\graphicspath{{Graphics/}}

\def\be{\begin{equation}}
\def\ee{\end{equation}}
\def\bea{\begin{eqnarray}}
\def\eea{\end{eqnarray}}

\definecolor{vividviolet}{rgb}{0.62, 0.0, 1.0}
\definecolor{amaranth}{rgb}{0.9, 0.17, 0.31}
\definecolor{palatinateblue}{rgb}{0.15, 0.23, 0.89}
\definecolor{brightpink}{rgb}{1.0, 0.0, 0.5}
\definecolor{cornflowerblue}{rgb}{0.39, 0.58, 0.93}
\definecolor{deepcarminepink}{rgb}{0.94, 0.19, 0.22}
\definecolor{radicalred}{rgb}{1.0, 0.21, 0.37}

\hypersetup{ linktoc=all,
    colorlinks, linkcolor={palatinateblue},
    citecolor={brightpink}, urlcolor={amaranth}
}

\begin{document}

\title{Entanglement degradation in regular and singular spacetimes}

\author{Orlando Luongo}
\email{orlando.luongo@unicam.it}
\affiliation{School of Science and Technology, University of Camerino, Via Madonna delle Carceri, Camerino, 62032, Italy.}
\affiliation{Istituto Nazionale di Fisica Nucleare (INFN), Sezione di Perugia, Perugia, 06123, Italy.}
\affiliation{Department of Nanoscale Science and Engineering, University at Albany SUNY, Albany, NY 12222, USA.}
\affiliation{INAF - Osservatorio Astronomico di Brera, Milano, Italy.}
\affiliation{Al-Farabi Kazakh National University, Al-Farabi av. 71, 050040 Almaty, Kazakhstan.}

\author{Stefano Mancini}
\email{stefano.mancini@unicam.it}
\affiliation{School of Science and Technology, University of Camerino, Via Madonna delle Carceri, Camerino, 62032, Italy.}
\affiliation{Istituto Nazionale di Fisica Nucleare (INFN), Sezione di Perugia, Perugia, 06123, Italy.}

\author{Sebastiano Tomasi}
\email{sebastiano.tomasi@unicam.it}
\affiliation{School of Science and Technology, University of Camerino, Via Madonna delle Carceri, Camerino, 62032, Italy.}
\affiliation{Istituto Nazionale di Fisica Nucleare (INFN), Sezione di Perugia, Perugia, 06123, Italy.}

\begin{abstract}
We study entanglement degradation near the horizons of regular, Reissner-Nordstr\"om, and Schwarzschild-de Sitter black holes, considering the Bardeen, Hayward, and generalized Hayward metrics as regular black holes.
To this end, we compute the entanglement negativity, $\mathcal{N}$, for two Unruh-like modes of a scalar field shared by Alice, who is inertial, and Rob, who hovers at a fractional offset $\rho$ outside the horizon of the backgrounds under consideration. For each geometry, we locally approximate the metric by a Rindler patch characterized by Rob's proper acceleration $a_0$. Because this Rindler approximation breaks down near the extremal limit, we also compute a near-extremal cutoff. Tracing over the inaccessible Rindler wedge yields a mixed Alice-Rob state, from which we evaluate $\mathcal{N}$ as a function of the mode frequency $\omega$ and the acceleration $a_0$. In all geometries considered, except for one, $\mathcal{N}$ increases monotonically with the parameter distinguishing that geometry form the Schwarzschild one. The exception is the Reissner-Nordstr\"om metric, for which $\mathcal{N}$ exhibits a shallow local minimum at a particular value of the charge. We also find that the Reissner-Nordstr\"om metric is the only background for which the negativity falls below that of the Schwarzschild case. Among all cases studied, the Schwarzschild-de Sitter spacetime provides the strongest protection of entanglement. Finally, across all backgrounds, high-frequency modes undergo less degradation than low-frequency modes. These results suggest that entanglement may serve as a useful probe for distinguishing Schwarzschild spacetime from other geometries.

\end{abstract}



\maketitle
\tableofcontents

\section{Introduction}

Understanding the main properties of entanglement in curved spacetimes represents one of the most important challenges in theoretical physics \cite{RevModPhys.76.93}, since it may help bridging quantum mechanics and gravity \cite{RyuTakayanagi2006HEE,VanRaamsdonk2010Building}. 

In this respect, strong-gravity domains appear as the natural scenarios in which one can investigate the consequences of entanglement in relativistic settings \cite{Martin-Martinez:2014gra,Belfiglio:2025cst}.

After the discovery of black hole imaging \cite{EHT2019M87Shadow,EHT2022SgrAShadow} and the recent results related to mimickers \cite{CardosoPani2019StatusReport,Bambi2025MimickersReview,Cunha2023LightRingInstability}, as alternatives to black holes, these realms can be used to test and quantify the influence of curvature on quantum correlations \cite{martinez0}. 

A well-known example of a horizon-induced quantum phenomenon is the the so-called \emph{Bekenstein-Hawking} temperature \cite{Bekenstein1973Entropy} and, more broadly, the corresponding predictions for thermal radiation emitted by black holes \cite{Hawking1975}. This phenomenon implies that particle pairs \cite{Israel1976Thermofield}, created near the event horizon, \emph{are entangled}\footnote{This presumes that one particle can escape the black hole as Hawking radiation, whereas its partner can fall behind the horizon; see, e.g., \cite{Wald1994}.}.

As a consequence of this phenomenon, one may expect that entanglement is severely degraded when one subsystem approaches a black hole horizon; see, e.g., \cite{Alsing2003}. 

Although entanglement degradation is not directly equivalent to the information loss problem \cite{Hawking1976,Page1993InfoRadiation,Almheiri2013Firewalls}, it provides a useful probe of how quantum correlations behave in strong gravitational fields. Nevertheless, if entanglement is irreversibly lost as black holes evaporate \cite{Mathur2009InfoParadox}, the evolution from a pure state to Hawking radiation would be nonunitary\footnote{Unitary evolution implies that a pure state remains pure. However, with entanglement degradation, an initially pure state evolves into a mixed one.}, dramatically violating the basic requirements of quantum mechanics. 

Generally, near a horizon, gravity induces an acceleration that can affect quantum entanglement, in agreement with the so-called \emph{Unruh effect} \cite{Crispino2008UnruhReview}, showing that an accelerating observer perceives the vacuum as a thermal bath \cite{Unruh1976}. The overall mechanism implies that acceleration, or, equivalently, the presence of a horizon, can degrade quantum correlations \cite{PhysRevA.92.022334}. Similarly, Hawking's analysis of black hole evaporation demonstrated particle creation at event horizons and raises important questions about information loss. 

These impressive developments open new avenues for studying entanglement in curved spacetimes while highlighting its inherent limitations. Research has shown that quantum teleportation fidelity decreases as acceleration increases \cite{LinChouHu2015Teleportation,Ge2005TeleportationSchwarzschild}. Furthermore, a state that appears maximally entangled in Alice's inertial frame is perceived as less entangled from Rob's perspective if he is uniformly accelerated. This degradation occurs because Rob, near the black hole horizon, is causally disconnected from a portion of spacetime and must trace over those inaccessible modes. Such findings confirmed that \emph{entanglement is observer-dependent: acceleration or the presence of a horizon leads to a loss of entanglement, essentially due to Unruh-Hawking thermal noise, which disrupts entanglement itself} \cite{Dai2015KillingEntanglement}.

Accordingly, entanglement in a variety of black hole spacetimes beyond the basic Schwarzschild case becomes particularly relevant. Examples include Schwarzschild-de Sitter (SdS) \cite{Gibbons1977}, which has both a black hole event horizon and a cosmological horizon, the Reissner-Nordstr\"om (RN) charged black hole \cite{Reissner1916,Nordstrom1918}, and regular black holes such as the Bardeen and Hayward models \cite{Bardeen1968,Hayward2006}, which modify the core geometry to avoid the central singularity. 

In principle, each of these spacetimes could alter the pattern of Hawking radiation and thus the way entanglement degrades \cite{PhysRevD.105.065007}. However, almost all detailed studies of entanglement degradation have been confined either to the Schwarzschild case or to its flat-space Rindler analogue. 

Motivated by the fact that entanglement degradation has not been studied in regular black holes, we explore the consequences of degradation in a specific class of mimickers, namely regular black holes \cite{Ansoldi2008RegularBHReview}. These objects have been introduced with the aim of inserting a de Sitter core that permits the removal of the singularity, in agreement with quantum mechanics \cite{Lan:2023cvz}. Recent developments show that regular black holes may follow the same thermodynamics as black holes, exhibiting an analogous area law \cite{Belfiglio:2024qsa,Belfiglio:2025hzo}, quantum properties \cite{Frolov:2016pav,Binetti:2022xdi,DelPiano:2023fiw}, repulsive effects \cite{LuongoQuevedo2023Repulsive}, and so on. Accordingly, the need to verify how entanglement degrades in charged or, more broadly, regular black hole configurations, as well as to perform a systematic comparison among these different background metrics, turns out to be quite important. In this respect, we carry out a systematic comparison of entanglement degradation in \emph{regular} and \emph{charged black hole backgrounds}, considering a simple setup in which two modes of a quantum scalar field are initially prepared in a maximally entangled state. As an initial setting, one mode is localized near the black hole horizon, potentially interacting with it, while the other remains in an inertial frame. By computing the remaining entanglement between the modes in the accelerated frame, quantified by the \emph{entanglement negativity}, we track how much entanglement is lost as a function of the black hole parameters. Our treatment appears model-dependent, as the free parameters of a given metric influence the degradation itself. We determine how model-dependent the entanglement measurements \emph{near the horizons} of different spacetimes are, checking whether the regularity of some solutions may be responsible for our outcomes. We apply this method to the Hayward metric and its generalization, and also to the Bardeen regular solution. Since these metrics depend on a cosmological constant and a magnetic charge, respectively, we compare our findings with the black hole counterparts, namely the SdS and RN spacetimes. In doing so, we compute the entanglement negativity for each metric and compare our findings with previous results, among them those of Refs. \cite{Fuentes2005,martinez1}. We found that entanglement degradation is not merely a consequence of the presence of a horizon; it is controlled by the specific parameters of the black hole metric. Several key insights emerge from our comparative analysis. First, the negativity can discriminate between Schwarzschild and regular or charged cores, as well as among different regular and charged backgrounds. In the RN spacetime, we observed a non-monotonic behavior. It is the only case in which the degradation is lower than in the Schwarzschild background. We also found that, in the SdS metric, the cosmological constant is particularly effective at restoring entanglement near the extremal configuration. Finally, across all backgrounds, high-frequency modes consistently exhibit less degradation than low-frequency modes.

The paper is organized as follows. In Sect. \ref{sec:local_approx}, we describe our approximation scheme for the background geometry. Afterwards, in Sect. \ref{sec:ent_degrad_theory}, we present the entanglement measure and explain how to compute it. The comparative results for entanglement in the various spacetimes are reported in Sect. \ref{sec:results}. We then discuss the implications of these findings for relativistic quantum information in Sect. \ref{sec:impact} and present the final outlook in Sect. \ref{sec:discussion}. Throughout this work, we employ natural units ($G=c=1$).


\section{The local metric approximation}\label{sec:local_approx}

We start from the fact that, in a spherically symmetric, static black hole spacetime, an observer standing at a distance $r_0$ from the black hole horizon can be effectively described by a Rindler metric with the observer's proper acceleration being the Rindler acceleration. 

A spherically symmetric metric can be cast into Rindler form with the only assumption that the observer does not sit exactly on the horizon.

\subsection{Metrics in Schwarzschild coordinates}\label{sec:schwarzschild-like}

Consider the line element for a spherically symmetric, static metric, in Schwarzschild coordinates, namely when the metric functions are the same, say
\begin{equation}\label{eq:schwarzshild_like_metric}
    \mathrm{d} s^2 = -f(r)\,\mathrm{d}t^2 + \frac{\mathrm{d}r^2}{f(r)} + r^2\,\mathrm{d}\Omega^2,
\end{equation}
For the subsequent calculations, we can ignore the angular part, setting  $\mathrm{d}\Omega^2 = 0$, reinstating it at the end.

The only prescription we need is that the underlying metric possesses an outer horizon, namely $r_h$.

Let us now consider an observer at a fixed radius, $r = r_0$, with $r_0>r_h$, rewriting Eq. \eqref{eq:schwarzshild_like_metric} as
\begin{equation}\label{sionz}
    \mathrm{d}\tau = \sqrt{f_0}\,\mathrm{d}t, \quad \text{with} \quad f_0 = f(r_0),
\end{equation}
where $\tau$ turns out to be the observer's proper time. For practical purposes, following a standard nomenclature, we may indicate the static observer as \emph{Rob}.

Thus, Eq. \eqref{eq:schwarzshild_like_metric} can be recast in function of Rob's proper time as
\begin{equation}
    \mathrm{d} s^2 = -\frac{f(r)}{f_0}\,\mathrm{d}\tau^2 + \frac{\mathrm{d}r^2}{f(r)}.
\end{equation}
Expanding up to first order around $r_0$ yields
\begin{equation}
f(r) \stackrel{r\rightarrow r_0}{\approx} f(r_0)+\left.\frac{\partial f}{\partial r}\right|_{r_0}(r - r_0).    
\end{equation}
Hence, defining
$f_0^{\prime} = \left.\frac{\partial f}{\partial r}\right|_{r_0} \quad \text{and} \quad \epsilon = r - r_0$, the metric can be approximated at first order as
\begin{equation}\label{metrexp}
    \mathrm{d} s^2 \approx -\frac{f_0+f_0^{\prime}\,\epsilon}{f_0}\,\mathrm{d}\tau^2 + \frac{\mathrm{d}\epsilon^2}{f_0+f_0^{\prime}\,\epsilon}.
\end{equation}
To obtain Rindler-like coordinates, we introduce the proper radial distance
\begin{equation}
    \mathrm{d}R=\frac{\mathrm{d}\epsilon}{\sqrt{f_0+f_0'\epsilon}} .
\end{equation}
Integrating and fixing the integration constant such that $R=R_0$ at
$\epsilon=0$, we obtain
\begin{equation}
    R-R_0=
\frac{2}{f_0'}
\left(
\sqrt{f_0+f_0'\epsilon}-\sqrt{f_0}
\right).
\end{equation}
Since $R_0$ only fixes the origin of the radial coordinate, we can shift
$R\rightarrow R-R_0$ and set $R_0=0$ without loss of generality.
The previous relation can then be inverted as $
\sqrt{f_0+f_0'\epsilon}
=
\sqrt{f_0}+\frac{f_0'}{2}R$. Thus, substituting into the metric yields
$f_0+f_0'\epsilon=
\left(\sqrt{f_0}+\frac{f_0'}{2}R\right)^2$, so that the line element becomes
\begin{equation}
\mathrm{d}s^2
\approx
-
\left(
1+\frac{f_0'}{2\sqrt{f_0}}R
\right)^2
\mathrm{d}\tau^2
+
\mathrm{d}R^2
+
r^2(R)\mathrm{d}\Omega^2 .
\end{equation}
This corresponds to the Rindler metric written locally around the static
observer Rob, whose worldline is located at $R=0$.
Hence, the metric can be expressed in terms of Rob's proper acceleration,
denoted by $a_0$, as
\begin{equation}
\mathrm{d}s^2
\approx
-
\left(
1+a_0R
\right)^2
\mathrm{d}\tau^2
+
\mathrm{d}R^2
+
r^2(R)\mathrm{d}\Omega^2 ,
\end{equation}
where the proper acceleration reads
\begin{equation}
a_0=\frac{f_0'}{2\sqrt{f_0}} .
\end{equation}
Since our treatment is only approximate, we cannot allow $r_0$ to take arbitrary values; accordingly, we require $r_0$ to lie very close to the black hole's outer event horizon. Thus, we set
\begin{equation}
    r_0 = r_h + \delta,
\end{equation}
where $\delta > 0$ is of the same order as $\epsilon$, that is, $\delta = \mathcal{O}(\epsilon)$. Hence, to leading order, we have $\delta\epsilon \approx 0$. Therefore
\begin{equation}
    f_0 \approx f'_h\,\delta\,,\qquad f'_0 \approx f'_h,
\end{equation}
and thus
\begin{equation}\label{eq:proper_acc_delta}
    a_0 \approx \sqrt{\frac{\kappa}{2\delta}},
\end{equation}
where $\kappa=f'_h/2$ is the \emph{surface gravity}. In this limit, the proper distance to the Rindler horizon is the same as the proper distance to the black hole horizon. Hence, the Rindler horizon provides a good approximation to the black hole horizon, as long as we consider sufficiently localized phenomena such that they are not affected by the spherical curvature of the horizon \cite{Crispino:2007eb}. 

As it will appear clear later, in our paper the above condition  means to consider \emph{field modes with wavelength much shorter than the horizon radius}.

\subsection{General spherically symmetric metric}

We now extend the results of the previous section by generalizing Eq. \eqref{eq:schwarzshild_like_metric} to
\begin{equation}\label{eq:general_sph_symm_metric}
    \mathrm{d}s^2 = -f(r)\,\mathrm{d}t^2 + \frac{\mathrm{d}r^2}{g(r)} + r^2\,\mathrm{d}\Omega^2,
\end{equation}
where $f(r)$ and $g(r)$ are, \emph{a priori}, distinct functions.

In this coordinate gauge, a hypersurface $r=r_h$ is null when $g(r_h)=0$. It is a Killing horizon of the static Killing field $\partial_t$ when $f(r_h)=0$. For a non-extremal static black hole, one typically requires both conditions, $f(r_h)=g(r_h)=0$, with simple zeros at $r_h$. The second condition ensures that the horizon is an \emph{infinite-redshift surface}, namely that $\partial_t$ becomes null there, so that no static observer can remain at $r=r_h$. The first condition guarantees that the hypersurface is lightlike and, together with the appropriate near-horizon behavior of $f$ and $g$, leads to a finite surface gravity.

We now adapt the previous construction to this more general setting and show that a local Rindler patch again emerges for the metric \eqref{eq:general_sph_symm_metric}.

As before, let $r_h$ denote the relevant horizon. Consider Rob as a static observer located at $r_0>r_h$. Restricting to the radial sector and using Rob's proper time as the time coordinate, the metric (deprived of its angular part) becomes
\begin{equation}
    \mathrm{d}s^2 = -\frac{f(r)}{f_0}\,\mathrm{d}\tau^2 + \frac{\mathrm{d}r^2}{g(r)},
\end{equation}
where $f_0 \equiv f(r_0)$, following the same prescription as in Eq. \eqref{sionz}. Near Rob's position, let $\epsilon \equiv r-r_0$ and expand the metric functions to first order, analogously to Eq. \eqref{metrexp}. This gives
\begin{equation}
    \mathrm{d}s^2 \approx -\left(1+ \frac{f_0^{\prime}}{f_0}\,\epsilon\right)\,\mathrm{d}\tau^2 + \frac{1}{g_0 + g_0^{\prime}\epsilon}\,\mathrm{d}\epsilon^2,
\end{equation}
where $g_0 \equiv g(r_0)$.

To cast the radial part in Rindler-like form, we introduce the proper distance $R$ through
\begin{equation}
    \mathrm{d}R = \frac{\mathrm{d}\epsilon}{\sqrt{g_0+g_0^{\prime}\epsilon}}\,,
\end{equation}
so that, choosing the integration constant such that $R=R_0$ at $\epsilon=0$,
\begin{equation}\label{eq:radial_variable_change}
    R-R_0 = \frac{2}{g_0^{\prime}}\left(\sqrt{g_0^{\prime}\epsilon + g_0}-\sqrt{g_0}\right)\approx \frac{\epsilon}{\sqrt{g_0}}\,.
\end{equation}
Differently from the previous section, we keep only the first-order term in $\epsilon$, since higher-order terms would generate curvature corrections and the metric would no longer be exactly of Rindler form. Substituting $\epsilon \approx \sqrt{g_0}(R-R_0)$ into the metric, we obtain
\begin{equation}
        \mathrm{d}s^2 \approx -\left(1+ \frac{f_0^{\prime}}{f_0}\sqrt{g_0}( R-R_0 )\right)\,\mathrm{d}\tau^2 + \mathrm{d}R^2.
\end{equation}
Since the proper acceleration of a static observer at $r_0$ is
\begin{equation}
    a_0 = \frac{1}{2}\frac{f'_0}{f_0}\sqrt{g_0},
\end{equation}
the metric can be rewritten as
\begin{equation}\label{eq:general_apprx_metric}
        \mathrm{d}s^2 \approx -\left[1+ 2a_0( R-R_0 )\right]\,\mathrm{d}\tau^2 + \mathrm{d}R^2,
\end{equation}
which is precisely the first-order expansion\footnote{This approximation is valid only in a sufficiently small neighborhood of $R_0$.} of the Rindler metric, since $\left[1 + a_0( R-R_0 )\right]^2\approx 1 + 2a_0( R-R_0 )$.

Hence, also in the general spherically symmetric case, the metric can be expanded around Rob's worldline in Rindler form, with Rob's proper acceleration playing the role of the Rindler acceleration \cite{Luongo:2025dsu}.

We are again interested in the regime in which $r_0$ lies very close to the horizon $r_h$. In analogy with Eq. \eqref{eq:proper_acc_delta}, let
\begin{equation}
    r_0=r_h+\delta,
\end{equation}
with $\delta>0$ small. Using the near-horizon expansions $f(r_0)\approx f'_h\delta$, $g(r_0)\approx g'_h\delta$, and $f'_0\approx f'_h$, we then find
\begin{equation}\label{a0fg1}
a_0 \approx\sqrt{\frac{g_h'}{4\delta}}.
\end{equation}
Introducing the standard expression
\begin{equation}
\kappa=\frac{1}{2}\sqrt{f'_hg'_h},
\end{equation}
for the surface gravity of a non-extremal horizon in the metric \eqref{eq:general_sph_symm_metric} \cite{KunduriLucietti2013NearHorizonReview}, Eq. \eqref{a0fg1} can be rewritten as
\begin{equation}
a_0=\frac{\kappa}{\sqrt{f_h'\,\delta}}. 
\end{equation}
For completeness, this construction also breaks down when $\kappa=0$, namely for extremal configurations. In that case, the derivation above is no longer valid because it assumes simple zeros of $f(r)$ and $g(r)$ at the horizon, so the linear near-horizon Taylor expansions used here must be replaced by a different analysis.

\subsection{Extremal black hole case}\label{sec:extremal_bound}

It is interesting to note that our discussion, although admittedly approximate, remains applicable provided that we do not consider extremal black holes. In that case, $f_h'$ also vanishes at the horizon and, therefore, the strategy described above cannot be employed \cite{KunduriLucietti2013NearHorizonReview}. In fact, to second order in the near-horizon expansion, the metric cannot be cast into a Rindler form, unlike in the non-extremal case. Rather, it reduces to a two-dimensional Anti-de Sitter space \cite{BardeenHorowitz1999ExtremeKerrThroat}. This regime lies beyond the main scope of the present paper.

Phrased differently, Eq.~(\ref{eq:proper_acc_delta}) is not valid for extremal black holes, since in that case the proper acceleration approaches $a_0 \approx \sqrt{f_h''/2}$. Therefore, a naive application of Eq.~(\ref{eq:proper_acc_delta}) to the extremal case would incorrectly predict $a_0=0$. One must then ask how \emph{near-extremal} the horizon may be before the Rindler approximation breaks down. This point is essential for quantifying $\delta$ even in the non-extremal case considered in this work. Expanding $f(r)$ about $r_0$ up to second order, one obtains $f(r)\approx f_0 + f_0'\,\epsilon + \frac{1}{2}f_0''\,\epsilon^2$. Accordingly, our approximation remains valid only if the quadratic term is negligible compared with the linear one, i.e., $\frac{1}{2}\,\lvert f_0''\,\epsilon\rvert \ll \lvert f_0'\rvert$. If $r_0$ is close to the horizon, we may replace the derivatives evaluated at $r_0$ by their horizon values, yielding
\begin{equation}\label{passag}
\epsilon \ll 2\frac{\lvert f_h'\rvert}{\lvert f_h''\rvert}
=4\frac{\lvert \kappa\rvert}{\lvert f_h''\rvert}.
\end{equation}
Defining $\alpha=\epsilon/r_h$, and using $r=r_0+\epsilon$ with $r_0\simeq r_h$, we have at leading order $\alpha \approx r/r_h-1$. Dividing both sides of Eq.~\eqref{passag} by $r_h$, we obtain
\begin{equation}\label{dist}
\lvert \alpha\rvert \ll 2\frac{\lvert f_h'\rvert}{r_h\lvert f_h''\rvert}.
\end{equation}
For practical purposes, we adopt a conservative numerical stopping criterion well within the regime of validity of Eq.~\eqref{dist}, and we terminate the numerical calculations when
\begin{equation}\label{proc}
\lvert \alpha\rvert \leq \frac{1}{100}\,\frac{2\lvert f_h'\rvert}{r_h\lvert f_h''\rvert}.
\end{equation}
That is, we stop the computation when the ratio between $\lvert \alpha\rvert$ and the right-hand side of Eq.~\eqref{dist} is of order $10^{-2}$, as specified in Eq.~\eqref{proc}.

Since in our subsequent simulations $\alpha \approx 10^{-2}$, this criterion reduces, up to factors of order one, to
\begin{equation}\label{eq:rindler_validity_condition}
r_h \leq \frac{2\lvert f_h'\rvert}{\lvert f_h''\rvert},
\end{equation}
thereby effectively controlling the deviation from extremality for metrics that admit such a limit. For a more general spherically symmetric metric, there would be two corresponding conditions, one for each metric function, and both would need to be satisfied simultaneously.

The extremal case is particularly important for our computation. For example, later in the paper we will consider a charged black hole which, as is well known, becomes extremal when the charge equals the mass.

\section{Entanglement degradation}\label{sec:ent_degrad_theory}

In this section, we describe how to compute the degradation of entanglement for a quantum field mode when one observer is uniformly accelerated and how this near-horizon analysis can be adapted locally to a black hole spacetime. Rindler coordinates are coordinates on a wedge of Minkowski spacetime adapted to uniformly accelerated observers. Owing to the Unruh effect, such an observer perceives the Minkowski vacuum as thermal. Therefore, when an inertial observer (Alice) and an accelerated observer (Rob) share a maximally entangled state, the entanglement accessible to Rob is reduced because he must trace over field modes beyond his Rindler horizon.

\subsection{Coordinates}
\begin{figure*}[t]
    \centering
    \begin{tikzpicture}[scale=1]
      \coordinate (iP)  at ( 0,  4);   
      \coordinate (iM)  at ( 0, -4);   
      \coordinate (ioL) at (-4,  0);   
      \coordinate (ioR) at ( 4,  0);   
      \coordinate (O)   at ( 0,  0);   
      \coordinate (HUR) at ( 2,  2);
      \coordinate (HUL) at (-2,  2);
      \coordinate (HDR) at ( 2, -2);
      \coordinate (HDL) at (-2, -2);
      \draw[line width=0.8pt] (iP) -- (ioR) -- (iM) -- (ioL) -- cycle;
      \draw[line width=0.6pt] (O) -- (iP);
      \draw[line width=0.6pt] (O) -- (ioR);
      \draw[line width=0.6pt] (O) -- (iM);
      \draw[line width=0.6pt] (O) -- (ioL);
      \draw[line width=0.9pt] (O) -- (HUR);
      \draw[line width=0.9pt] (O) -- (HUL);
      \draw[line width=0.9pt] (O) -- (HDR);
      \draw[line width=0.9pt] (O) -- (HDL);
      \draw[thick,smooth]
        plot coordinates {(2,2) (1.4,1) (1.1,0) (1.4,-1) (2,-2)};
      \draw[black,thick]
        (0,-4) to[out=70,in=-70] (0,4);
      \node[above=2pt]               at ( 0,  4) {$i^{+}$};
      \node[below=2pt]               at ( 0, -4) {$i^{-}$};
      \node[anchor=east,xshift=-2pt] at (-4,  0) {$i^{0}$};
      \node[anchor=west,xshift= 2pt] at ( 4,  0) {$i^{0}$};
      \node[anchor=south east, xshift=-2pt, yshift= 2pt] at (-2,  2) {$\mathcal{I}^{+}$};
      \node[anchor=south west, xshift= 2pt, yshift= 2pt] at ( 2,  2) {$\mathcal{I}^{+}$};
      \node[anchor=north east, xshift=-2pt, yshift=-2pt] at (-2, -2) {$\mathcal{I}^{-}$};
      \node[anchor=north west, xshift= 2pt, yshift=-2pt] at ( 2, -2) {$\mathcal{I}^{-}$};
      \node[rotate=-45, anchor=south west, xshift= 1pt, yshift= 1pt]
        at ( 1.6,  1.6) {$\mathcal{H}^{+}$};
      \node[rotate= 45, anchor=south east, xshift=-1pt, yshift= 1pt]
        at (-1.6,  1.6) {$\mathcal{H}^{-}$};
      \node[rotate= 45, anchor=north west, xshift= 1pt, yshift=-1pt]
        at ( 1.6, -1.6) {$\mathcal{H}^{-}$};
      \node[rotate=-45, anchor=north east, xshift=-1pt, yshift=-1pt]
        at (-1.6, -1.6) {$\mathcal{H}^{+}$};
      \node[anchor=west,  xshift= 2pt] at ( 2,  0.25) {I};
      \node[anchor=south, yshift= 2pt] at ( -0.25,  2.5) {II};
      \node[anchor=north, yshift=-2pt] at ( -0.25, -2.5) {III};
      \node[anchor=east,  xshift=-2pt] at (-2,  0.25) {IV};
      \node[above=1pt] at ( 1.5,  0.4) {R};
      \node[below=1pt] at ( 0.7, -1.9) {A};
    \end{tikzpicture}
    \caption{Penrose (conformal) diagram of Minkowski spacetime showing a uniformly accelerated (Rindler) observer $R$ and an inertial observer $A$. The null lines $x=\pm t$ partition spacetime into four connected regions (Rindler wedges). For observers with constant proper acceleration in the right wedge ($x>|t|$), these null boundaries are acceleration (Rindler) horizons: the future horizon $\mathcal{H}^+$ is $t=x$ and the past horizon $\mathcal{H}^-$ is $t=-x$. The worldline of $R$ lies completely within the right (first) Rindler wedge and is asymptotic to $\mathcal{H}^\pm$.}
    \label{fig:rindler_diagram}
\end{figure*}
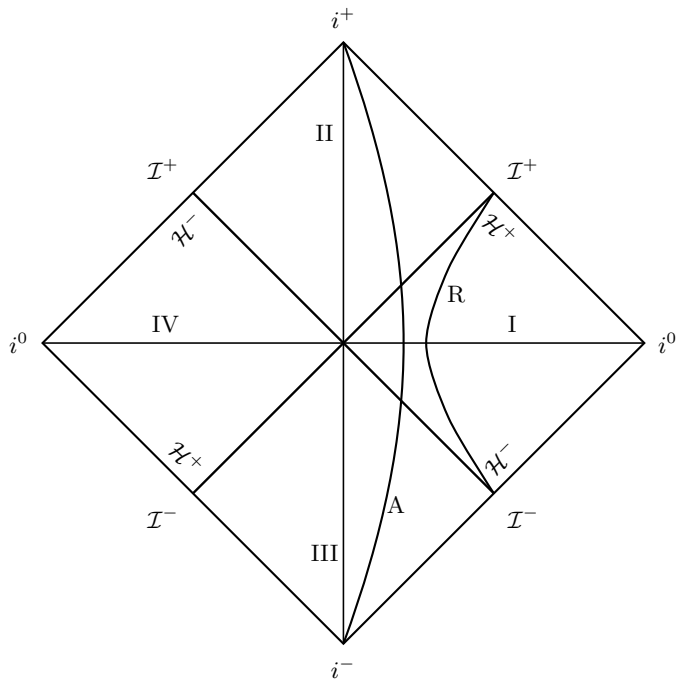

To set the stage, we begin by recalling the transformation from Minkowski coordinates $(\hat{t}, \hat{x}, \hat{y}, \hat{z})$ to Rindler coordinates $(t,x,y,z)$ for an observer uniformly accelerated along the $z$-axis. In region I of Rindler space, the transformation reads
\begin{align}
a_0 \hat{t} &= e^{a_0 z} \sinh(a_0 t), \\
a_0 \hat{z} &= e^{a_0 z} \cosh(a_0 t), \\
\hat{x} &= x,\quad \hat{y} = y,
\end{align}
where $a_0$ sets the acceleration scale. For the reference trajectory $z=0$, which we identify with Rob's worldline, the proper acceleration is $a_0$. This coordinate change leads to the appearance of a Rindler
horizon, beyond which Rob has no causal contact. The conformal diagram is shown in Fig.~\ref{fig:rindler_diagram}.

In Minkowski spacetime, a field can be quantized canonically. In this context, one has well-defined positive- and negative-frequency mode functions which, for a massless scalar field, are solutions of the wave equation. We choose a complete set of these modes normalized with respect to the Klein-Gordon inner product. Each mode is also an eigenfunction of the Minkowski timelike Killing vector $\partial_{\hat t}$ and therefore carries a conserved Minkowski energy.

A second relevant Killing field is the Lorentz-boost generator, which becomes $\partial_t$ in Rindler coordinates. This vector is timelike only within the Rindler wedges, where it defines positive frequency with respect to Rindler time. One can therefore define a complete set of Rindler mode functions that are eigenfunctions of $\partial_t$. The Minkowski and Rindler mode bases are not independent: any mode in one basis may be expanded in terms of the other. The coefficients in this expansion are the Bogoliubov coefficients, and they induce the corresponding map between Minkowski and Rindler ladder operators in the quantized theory.

Both choices of positive frequency induce Fock-space representations, with corresponding annihilation operators and vacuum states. In the Rindler description, the vacuum is the product state annihilated by the annihilation operators in wedges I and IV. In the field-theory setting, the Rindler and Minkowski vacuum states are not unitarily equivalent \cite{Fulling1973NonUniqueness}. Consequently, what appears as vacuum in Minkowski spacetime contains particles when viewed by a Rindler observer.

\subsection{Bogoliubov transformation}
The \emph{Bogoliubov transformation} relates the Minkowski mode operators to those defined in the Rindler wedges. For a scalar field, one can show that the relevant Minkowski annihilation\footnote{Here, by ``Minkowski operator'' we mean an operator that creates or annihilates Unruh-like modes, not ordinary monochromatic Minkowski plane-wave modes. It nevertheless annihilates the Minkowski vacuum.} operator $a_{\omega}$ can be written as
\begin{equation}
a_{\omega} = \cosh (p)\, a_{\omega,I} - \sinh (p)\, a^\dagger_{\omega,IV},
\label{eq:bogoliubov}
\end{equation}
where $a_{\omega,I}$ is the annihilation operator in the right Rindler wedge, $a^\dagger_{\omega,IV}$ is the creation operator in the left wedge, and the parameter $p$ is determined by
\begin{equation}
\tanh p = \exp\Bigl(-\frac{\pi \omega}{a_0}\Bigr),
\label{eq:rs}
\end{equation}
with $\omega$ the Rindler mode frequency. This transformation implies that the Minkowski vacuum state is a two-mode squeezed state in the Rindler Fock space:
\begin{equation}
|0\rangle_M = \frac{1}{\cosh p}\sum_{n=0}^{\infty} (\tanh p)^n\, |n_{\omega}\rangle_I \otimes |n_{\omega}\rangle_{IV}.
\label{eq:mink_vacuum}
\end{equation}
By acting with the corresponding Minkowski creation operator on the Minkowski vacuum, one also obtains the Minkowski one-particle state in terms of Rindler modes. Using Eq.~\eqref{eq:bogoliubov}, one finds
\begin{equation}\label{eq:one_particle_state}
|1_{\omega}\rangle_M
=\frac{1}{\cosh^2 p}
\sum_{n=0}^{\infty} (\tanh p)^n\,\sqrt{n+1}\;
\bigl|{(n+1)}_{\omega}\bigr\rangle_I \otimes \bigl|n_{\omega}\bigr\rangle_{IV}.
\end{equation}

\subsection{Entanglement degradation}
In the entanglement setup, Alice is inertial, while Rob is held static at fixed altitude close to a black hole horizon. Alice and Rob initially meet far from the black hole, where spacetime is approximately Minkowskian. There they prepare a maximally entangled state such as
\begin{equation}
|\Psi\rangle = \frac{1}{\sqrt{2}} \Bigl(|0\rangle_A |0\rangle_R + |1\rangle_A |1_{\omega}\rangle_R\Bigr),
\label{eq:entangled_state}
\end{equation}
where $|0\rangle_A$ and $|1\rangle_A$ denote Alice's qubit basis states, $|0\rangle_R \equiv |0\rangle_M$ is the Minkowski vacuum of Rob's mode, and $|1_{\omega}\rangle_R$ is the corresponding one-particle excitation at frequency $\omega$. Rob then travels toward the black hole and, close to the event horizon, fires his engines to hover outside it. In a local near-horizon description, the exterior geometry is approximately Rindler, so Rob rewrites his field excitations in the Rindler basis as Eq.(\ref{eq:one_particle_state}):
\begin{equation}
    \begin{aligned}
        |\Psi\rangle
        &=\frac{1}{\sqrt{2}\,\cosh^{2}p}\sum_{n=0}^{\infty}(\tanh p)^{n}
        \Bigl[\cosh p\,|0\rangle_A\,|n_{\omega}\rangle_I\,|n_{\omega}\rangle_{IV}
        \\&+\sqrt{n+1}\,|1\rangle_A\,|(n+1)_{\omega}\rangle_I\,|n_{\omega}\rangle_{IV}
        \Bigr].
    \end{aligned}
\end{equation}
As a result, once Rob traces over the inaccessible modes, the originally pure state becomes mixed and the accessible entanglement is reduced. In the near-horizon approximation, the inaccessible modes are those in region IV, which model degrees of freedom beyond the horizon.

To characterize what Rob experiences, we therefore trace over the region-IV modes. The resulting density matrix is
\begin{equation}
\rho_{AR} = \operatorname{Tr}_{IV} \bigl(|\Psi\rangle\langle\Psi|\bigr).
\label{eq:partial_trace}
\end{equation}
More explicitly we have
\begin{equation}\label{eq:ar_density_matrix}
\begin{aligned}
\rho_{AR}
=&
\sum_{n=0}^{\infty}
\frac{\tanh^{2n} p}{2\cosh^{2} p}
\Bigg[
|0,n\rangle\langle 0,n|
\\
&\quad+\frac{\sqrt{n+1}}{\cosh p}
\Bigl(
|0,n\rangle\langle 1,n\!+\!1|
+
|1,n\!+\!1\rangle\langle 0,n|
\Bigr)
\\
&\quad+\frac{n+1}{\cosh^{2} p}
|1,n\!+\!1\rangle\langle 1,n\!+\!1|
\Bigg],
\end{aligned}
\end{equation}
where we have dropped the wedge labels because only region-I states remain, with the shorthand notation
$|0,n\rangle \equiv |0\rangle_A |n_{\omega}\rangle_I$
and
$|1,n\!+\!1\rangle \equiv |1\rangle_A |(n+1)_{\omega}\rangle_I$.

Because the original pure state becomes mixed, the von Neumann entropy of a reduced state is no longer a faithful measure of entanglement. A more appropriate quantity is the \emph{negativity} \cite{VidalWerner2002Negativity} $\mathcal{N}$, defined for a bipartite state $\rho$ as
\begin{equation}
\mathcal{N} = -\sum_{\lambda_i < 0} \lambda_i,
\label{eq:negativity}
\end{equation}
where the $\lambda_i$ are the eigenvalues of the partial transpose of $\rho$ with respect to one subsystem. This quantity detects entanglement whenever the partial transpose has negative eigenvalues and is the standard measure in this setup. For a Bell-like pure state such as Eq.~\eqref{eq:entangled_state}, the negativity is maximal, with $\mathcal{N}=1/2$. By contrast, separable states have vanishing negativity. In our context, the initial state has maximal negativity. When Rob is placed near the horizon, the state becomes mixed and the negativity drops, signaling a loss of entanglement.

The negativity for the state Eq.(\ref{eq:ar_density_matrix}) results:
\begin{equation}
\label{eq:bosonic_negativity}
\begin{aligned}
\mathcal{N}_{AR}^{(s)}
&= \sum_{n=0}^{\infty}
\frac{\tanh^{2n} p}{4\,\cosh^{2}p}
\\
&\quad\times
\biggl[
\sqrt{\Bigl(\tfrac{n}{\sinh^{2}p} + \tanh^{2}p\Bigr)^{2}
+ \tfrac{4}{\cosh^{2}p}}
\\&-\Bigl(\tfrac{n}{\sinh^{2}p} + \tanh^{2}p\Bigr)
\biggr].
\end{aligned}
\end{equation}
In the inertial limit $a_0 \to 0$, one has $p \to 0$ and therefore $\tanh p \to 0$; the state remains nearly pure and maximal entanglement is recovered. As $a_0$ increases, $\tanh p$ increases, enhancing the weight of higher-occupation-number Rindler sectors and leading to a greater loss of entanglement.

In summary, by using the Bogoliubov transformations, Eqs.~\eqref{eq:bogoliubov} and \eqref{eq:rs}, to express Rob's Minkowski vacuum and one-particle states in the Rindler basis, one obtains a mixed Alice-Rob state after tracing over the inaccessible wedge IV. The resulting loss of entanglement is then quantified by the negativity, Eq.~\eqref{eq:bosonic_negativity}, which depends explicitly on Rob's proper acceleration scale $a_0$.

Throughout, we do \emph{not} rely on the single-mode approximation in the sense of monochromatic Minkowski plane waves. Instead, consistently with Eq.~\eqref{eq:bogoliubov}, the operators $a_{\omega}$ annihilate \emph{Unruh-like} modes, namely non-monochromatic superpositions of positive-frequency Minkowski plane waves chosen so that the Bogoliubov map is diagonal. With this choice, each $\omega$ couples only to the corresponding pair of Rindler modes $(\omega,I)$ and $(\omega,IV)$. We are therefore considering not a single mode in the Minkowski-plane-wave sense, but rather an effective \emph{single-Unruh-mode} description. More realistic detectors couple to wave packets and hence to superpositions of $\omega$, thus introducing additional bandwidth-dependent mixing. For sufficiently well-peaked packets, however, the qualitative dependence of $\mathcal{N}(\omega,a_0)$ on acceleration and frequency is expected to persist, with quantitative corrections controlled by the packet bandwidth and switching profile. In fact, Ref.~\cite{Bruschi2010} proved that there exists a controlled class of Minkowski wave packets for which the single-Unruh-frequency picture provides a good approximation.

\section{Spacetimes with horizons}

In this section, we briefly present the black hole models considered in our work, for which we compute the corresponding entanglement degradation.

Using the local approximation derived in Sect. \ref{sec:local_approx}, the spacetime in the vicinity of Rob can be described by a Rindler-like metric characterized by $a_0$. Accordingly, Rob's proper acceleration $a_0$ determines the entanglement degradation discussed in Sect. \ref{sec:ent_degrad_theory}. After briefly introducing each spacetime, we therefore provide the corresponding expression for $a_0$.

More generally, one may consider black hole spacetimes characterized by parameters in addition to mass.

In the standard classical picture, stationary black holes are characterized by a limited set of global charges, in accordance with the no-hair paradigm \cite{Herdeiro:2015waa}. Moreover, under the standard assumptions of the classical singularity theorems, gravitational collapse leads to spacetimes containing singularities.

More recently, however, regular black hole solutions have been proposed as alternatives to singular black hole geometries \cite{Volkov:2016ehx,Frolov:2016pav}. In many such models, the role of electric charge is replaced by topological charges, while cosmological-constant-like contributions can be interpreted as effective vacuum-energy terms \cite{Padmanabhan:2002ji,Belfiglio:2022qai,Belfiglio:2023rxb}.

Below, we distinguish between
\begin{itemize}
    \item[-] \emph{black holes},
    \begin{enumerate}
        \item with electric charge,
        \item with a cosmological constant,
    \end{enumerate}
    \item[-] \emph{regular black holes},
    \begin{enumerate}
        \item with magnetic charge,
        \item with a vacuum-energy contribution.
    \end{enumerate}
\end{itemize}
For each case discussed below, we comment on the existence of horizons and, consequently, on the applicability of the entanglement-degradation analysis.

\subsection{Black holes with charge: The Reissner-Nordstr\"om metric}\label{sec:charged_black_holes}

This spacetime describes a charged, non-rotating black hole in Schwarzschild coordinates, with line element as in Eq.~(\ref{eq:schwarzshild_like_metric}) and \cite{Giorgi:2019kjt}
\begin{equation}
f(r) = 1 - \frac{2M}{r} + \frac{Q^2}{r^2},
\end{equation}
where $M$ is the black hole mass and $Q$ its electric charge.

The horizon radii are
\begin{equation}
r_\pm = M \pm \sqrt{M^2 - Q^2},
\end{equation}
provided that $M^2 \ge Q^2$. For $M^2 > Q^2$, $r_+$ and $r_-$ are the outer and inner horizons, respectively; for $M^2 = Q^2$, the two horizons coincide; and for $M^2 < Q^2$, no horizon is present.

For a static observer (Rob) at fixed radius $r_0$ near the outer horizon, the local geometry is approximately Rindler-like, and the corresponding proper acceleration is
\begin{equation}
a_0 = \frac{\left|M r_0 - Q^2\right|}{r_0^2 \sqrt{r_0 (r_0 - 2M) + Q^2}},
\end{equation}
where $r_0 > r_+$.

\subsection{Regular black holes with magnetic charge: The Bardeen spacetime}

The Bardeen metric is a spherically symmetric, asymptotically flat spacetime. It has finite curvature invariants everywhere and is therefore regular. Physically, it can be interpreted as a regular black hole carrying a magnetic charge in a model of nonlinear electrodynamics \cite{AyonBeatoGarcia2000Bardeen}. It is defined by 
\begin{equation}
f(r) = 1 - \frac{2Mr^2}{\left(r^2+g^2\right)^{3/2}}.
\end{equation}
Here, $g$ is the magnetic charge parameter arising from the nonlinear-electrodynamics framework. In the limit $r \to \infty$ or $g\rightarrow 0$, the metric function approaches the Schwarzschild form. For a static observer at $r_0$, the proper acceleration is
\begin{equation}
a_0 = \frac{M \left|r_0^3 - 2 g^2 r_0\right|}{\left(g^2+r_0^2\right)^{5/2} \sqrt{1-\frac{2 M r_0^2}{\left(g^2+r_0^2\right)^{3/2}}}}.
\end{equation}

\subsection{Regular solutions with a de Sitter core: The Hayward metric}

The Hayward metric was proposed as a model for regular black holes incorporating a length scale $\ell$, often expected to be of the order of the Planck length, at which quantum-gravitational effects may become relevant \cite{Hayward:2005gi}. Such effects may eliminate the central curvature singularity. The metric function is
\begin{equation}
f(r) = 1 - \frac{2M r^2}{r^3 + 2M\ell^2}.
\end{equation}
In the asymptotic limit $r \to \infty$, the metric approaches the Schwarzschild solution, while for small $r$ it becomes de Sitter-like.

The parameter $\ell$ acts as a regulator and is useful for exploring the phenomenology of regular black holes and for developing insight into possible resolutions of black hole singularities.

For a static observer at $r_0$, the proper acceleration is
\begin{equation}
a_0 = \frac{M \left|r_0^4 - 4 \ell^2 M r_0\right|}{\left(2 \ell^2 M+r_0^3\right)^2 \sqrt{1-\frac{2 M r_0^2}{2 \ell^2 M+r_0^3}}}.
\end{equation}

\subsection{Regular solutions with a de Sitter core: The modified Hayward metric}

The modified Hayward metric is a static, spherically symmetric spacetime that extends the original Hayward regular black hole by incorporating both one-loop quantum corrections to the Newtonian potential and a finite time delay between an observer at infinity and one at the regular center \cite{LingWu2022ModifiedRegularBH}. The metric is of the form of Eq.(\ref{eq:general_sph_symm_metric}) with
\begin{equation}
    f(r)=G(r)\,F(r)\qquad g(r)=F(r),
\end{equation}
where
\begin{subequations}
\begin{align}
F(r) &= 1 - \frac{2M(r)}{r},\\
M(r) &= \frac{m r^3}{r^3+2m\ell^2},\\
G(r) &= 1 - \frac{\beta m\alpha}{\alpha r^3+\beta m},
\end{align}
\end{subequations}
with $m$ the ADM mass \cite{Arnowitt:1962hi,Carroll:1997ar}$,$ while $\alpha\in[0,1)$ parametrizes the time delay between the center and infinity and $\beta$ is a numerical constant, typically $\beta=41/(10\pi)$ \cite{BalartVagenas2014WEC}.

For $m>m_\star\equiv\tfrac{3\sqrt{3}}{4}\ell$, the equation $F(r_h)=0$ admits two horizons, denoted by $r_-$ and $r_+\equiv r_h$.

A static observer at $r_0>r_h$ has proper acceleration
\begin{equation}
a_0 \;=\; \frac{\bigl|\,G'(r_0)\,F(r_0)\;+\;G(r_0)\,F'(r_0)\bigr|}
     {2\,G(r_0)\,\sqrt{F(r_0)}}.
\end{equation}

For completeness, the one-loop correction enters only through the coupling $G(r)$, whereas in the limit $\alpha\to 0$ one recovers the standard Hayward spacetime exactly. Near the outer horizon, the metric again reduces locally to Rindler form.

\subsection{Black holes with a de Sitter phase: Schwarzschild-de Sitter}\label{sec:de_sitt_bh}

The SdS metric extends the Schwarzschild solution to include a positive cosmological constant $\Lambda$, which models a de Sitter phase \cite{Kottler:1918cxc}. Its line element is determined by
\begin{equation}
f(r) = 1 - \frac{2M}{r} - \frac{\Lambda}{3}\, r^2,
\end{equation}
with $\Lambda$ the cosmological constant. When $\Lambda \rightarrow 0$, the spacetime reduces to the Schwarzschild metric. Conversely, if $M \rightarrow 0$, the metric describes de Sitter spacetime in static coordinates\footnote{For an extended version of de Sitter space, see e.g. Ref.~\cite{Giambo:2023zmy}.}.

The SdS metric describes the gravitational field of a black hole embedded in an expanding universe. The cosmological constant modifies the asymptotic structure of the spacetime and, for suitable values of $M$ and $\Lambda$, introduces a cosmological horizon. It also affects the thermodynamic and causal structure. This metric is particularly relevant when studying the interplay between local gravitational dynamics and global cosmological expansion.

For a static observer at $r_0$, the proper acceleration is
\begin{equation}
a_0 = \frac{\left|3 M - \Lambda r_0^3\right|}{3 r_0^2 \sqrt{1 - \frac{2M}{r_0} - \frac{\Lambda}{3} r_0^2}}.
\end{equation}

When $0 < 9 \Lambda M^2 < 1$, the equation $f(r)=0$ admits two  positive roots. Denoting the smaller and larger ones by $r_-$ and $r_+$, respectively, the relevant horizon turns out to be $r_-$. Indeed, as $\Lambda \to 0$, one has $r_- \to 2M$, whereas the cosmological horizon behaves as $r_+ \approx \sqrt{\frac{3}{\Lambda}} \to \infty$. Hence, for small $\Lambda$, the static region accessible to stationary observers is
\begin{equation}\label{eq:cavity}
r_- < r < r_+,
\end{equation}
obtained as a consequence of the natural structure of the SdS metric. 

Nevertheless, in SdS spacetime, there are in general two horizons with different surface gravities and hence different temperatures. A static observer in the cavity Eq.(\ref{eq:cavity}) is not in equilibrium with a single thermal bath unless one imposes additional assumptions or restricts to a local analysis near one horizon only. Accordingly, choosing a small $\Lambda$ guarantees this to hold, as we will discuss even later in our findings.

\section{Numerical results on entanglement degradation}\label{sec:results}

We now quantify how quantum correlations are modified when a black hole is characterized not only by its mass, but also by additional external parameters.

To this end, we track the negativity $\mathcal N$, Eq.~(\ref{eq:bosonic_negativity}), of an initially maximally entangled state shared by \emph{Alice} (inertial) and \emph{Rob} (the usual static observer located at radius $r_0$ just outside the relevant horizon) as the background is varied from Schwarzschild to each of the metrics introduced above. We therefore compare the main departures from the standard Schwarzschild case and assess how the additional parameters affect entanglement degradation.

Conventionally, we fix the gravitational mass, $M=1$, and the dimensionless radial ratio $\rho=r_0/r_h=1.01$, leaving the remaining parameter free to vary.

Since the Rindler approximation breaks down near the extremal limit, we compute an upper bound on the parameter for each geometry using Eq.~(\ref{eq:rindler_validity_condition}). This bound ensures that the near-horizon approximation remains valid and that the corresponding results are physically meaningful.

\subsection{Results for the Reissner-Nordstr\"om spacetime}

\begin{figure}[H]
  \centering
  \includegraphics[width=\linewidth]{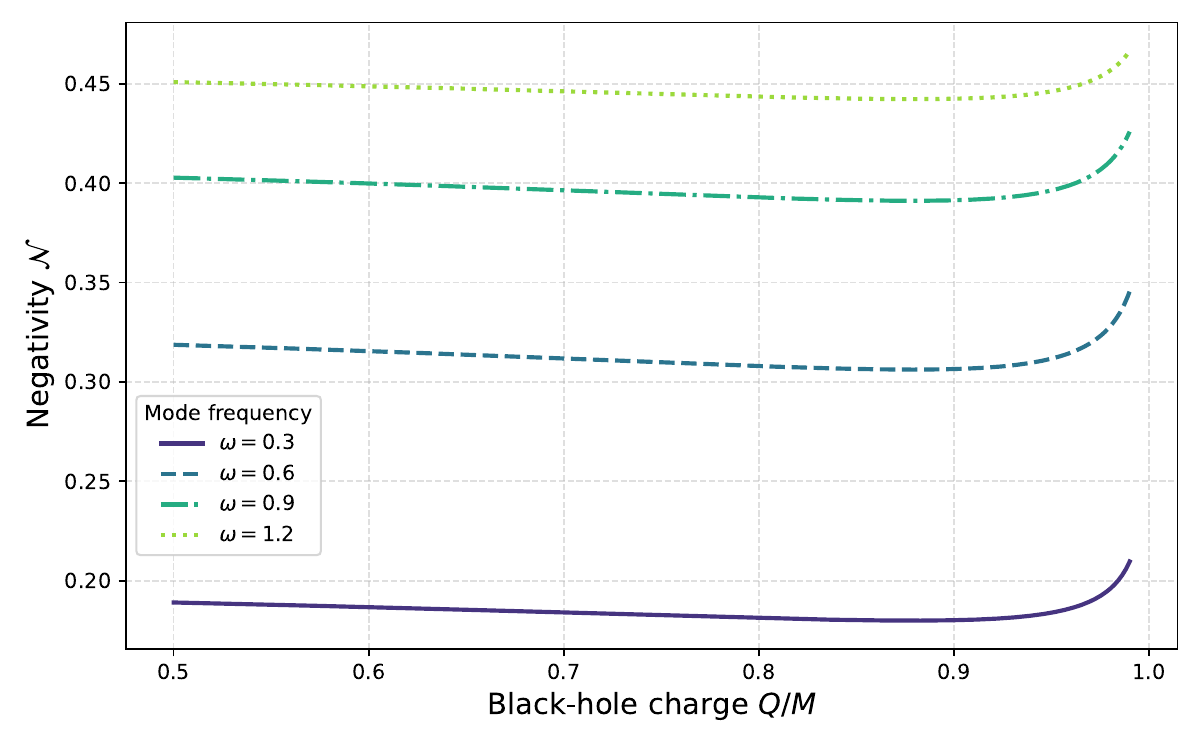}
  \caption{Negativity as a function of the charge-to-mass ratio $Q/M$ for several frequencies $\omega$ at fixed $\rho$. A  minimum in the negativity is present at $Q/M\approx0.879$.}
  \label{fig:nega_rn_charge}
\end{figure}

We begin with the RN metric. The varying parameter is the charge-to-mass ratio, $Q/M$, which controls the surface gravity and the associated near-horizon thermal bath. Accordingly, Fig.~\ref{fig:nega_rn_charge} summarizes our findings.

Starting from the neutral case, $Q=0$, and increasing $Q/M$, we observe a local minimum centered at $Q/M\approx0.879$. At this value of the charge, Rob's proper acceleration reaches a maximum; within the near-horizon description, the effect on entanglement of the local Unruh/Hawking temperature is also maximal, yielding the strongest degradation.

Near the upper bound, computed to be $Q/M\approx0.99$, the entanglement is better protected. Indeed, Fig.~\ref{fig:nega_rn_charge} shows a sharp increase in the negativity between the minimum and the upper bound.

Equation~\eqref{eq:q/m_min} gives the analytical location of the minimum and shows that it depends only on $\rho$:
\begin{equation}\label{eq:q/m_min}
    \begin{aligned}
        \left.\frac{Q}{M}\right|_{\mathrm{min}}(\rho)
  &=\frac{\sqrt{\,2\rho^{3}-2\rho^{4}-\frac32\left[1+s\right]
   +\rho^{2}\left[5+s\right]}}
  {1+\rho}\\
   s&=\sqrt{1+8\rho-4\rho^{2}-8\rho^{3}+4\rho^{4}}
    \end{aligned}
\end{equation}
For $\tfrac12\left(1+\sqrt{5-2\sqrt{3}}\right)\le\rho\le \tfrac12\left(1+\sqrt{5+2\sqrt{3}}\right)$, i.e., $1.12\,r_{+}\lesssim r_{0}\lesssim1.95\,r_{+}$, the radicand in Eq.~(\ref{eq:q/m_min}) is negative and no real minimum exists. Our choice of $\rho$ lies outside this interval; hence the dip seen in Fig.~\ref{fig:nega_rn_charge} is expected.

The Hawking temperature, $T_H=\kappa/(2\pi)$ is most naturally described in parametric form as
\begin{equation}
\left\{
\begin{aligned}
a_0 &= a_0(\lambda),\\
T_H &\approx \bigl[a_0(\lambda)\bigr]^2\,r_h(\lambda)\left(\frac{\rho-1}{\pi}\right),
\end{aligned}
\right.
\qquad \lambda=\frac{Q}{M}.
\end{equation}
The resulting curve is not single-valued for RN, when represented in the $(a_0,T_H)$ plane. The reason is that $a_0$, in this case, is not a monotonic function of the charge-to-mass ratio $Q/M$: two distinct values, $\lambda_1\neq\lambda_2$, may correspond to the same proper acceleration, $a_0(\lambda_1)=a_0(\lambda_2)$, while yielding different horizon radii, $r_h(\lambda_1)\neq r_h(\lambda_2)$. Consequently, the same value of $a_0$, and thus the same negativity, can be associated with two different Hawking temperatures. This is another peculiarity of the RN spacetime. As we will later show, the other geometries are simpler and allow for a one-to-one correspondence between the Hawking temperature and the entanglement degradation.

\begin{figure}[H]
    \centering
    \includegraphics[width=1\linewidth]{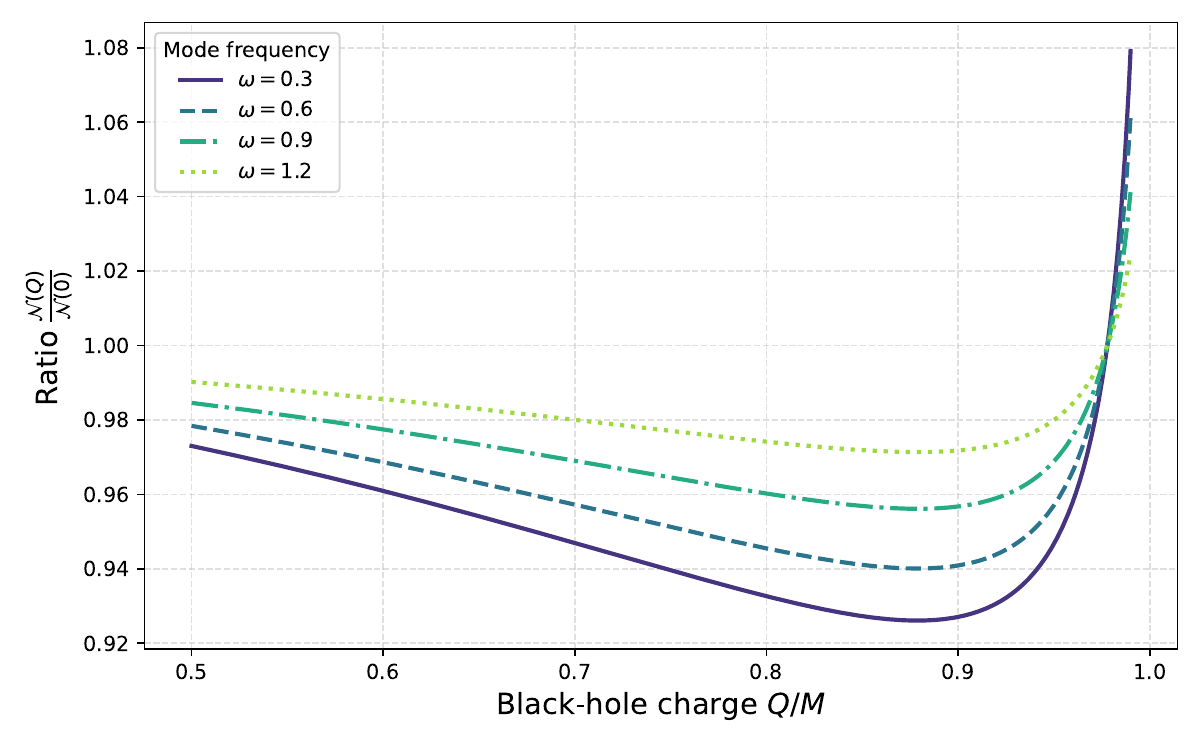}
    \caption{Negativity in the RN case, normalized to the corresponding Schwarzschild value at $Q/M=0$.}
    \label{fig:nega_rn_charge_perc}
\end{figure}

To facilitate comparison with the Schwarzschild case, Fig.~\ref{fig:nega_rn_charge_perc} shows the ratio of the negativity in the RN case to the corresponding Schwarzschild value. The plotted curves are obtained by normalizing each curve in Fig.~\ref{fig:nega_rn_charge} by its value at $Q/M=0$, which represents the Schwarzschild limit.

For small charges, the entanglement is more degraded than in the Schwarzschild case, until we reach\footnote{This can be recovered analytically by setting the Schwarzschild proper acceleration equal to the RN one. One gets $Q=\sqrt{1-s^2}$, where $s$ is the only physical solution of $(\alpha^2+3\alpha+2)s^3+(5\alpha^2+13\alpha+8)s^2+(11\alpha^2+25\alpha-2)s+\alpha(15\alpha-1)=0$, and we defined $\alpha = \rho -1$.} $Q/M\approx0.978$, where the negativity for all frequencies coincides with the Schwarzschild value. Beyond this point, increasing the charge helps preserve the entanglement.

\subsection{Results for the Bardeen spacetime}

\begin{figure}[H]
    \centering
    \includegraphics[width=\linewidth]{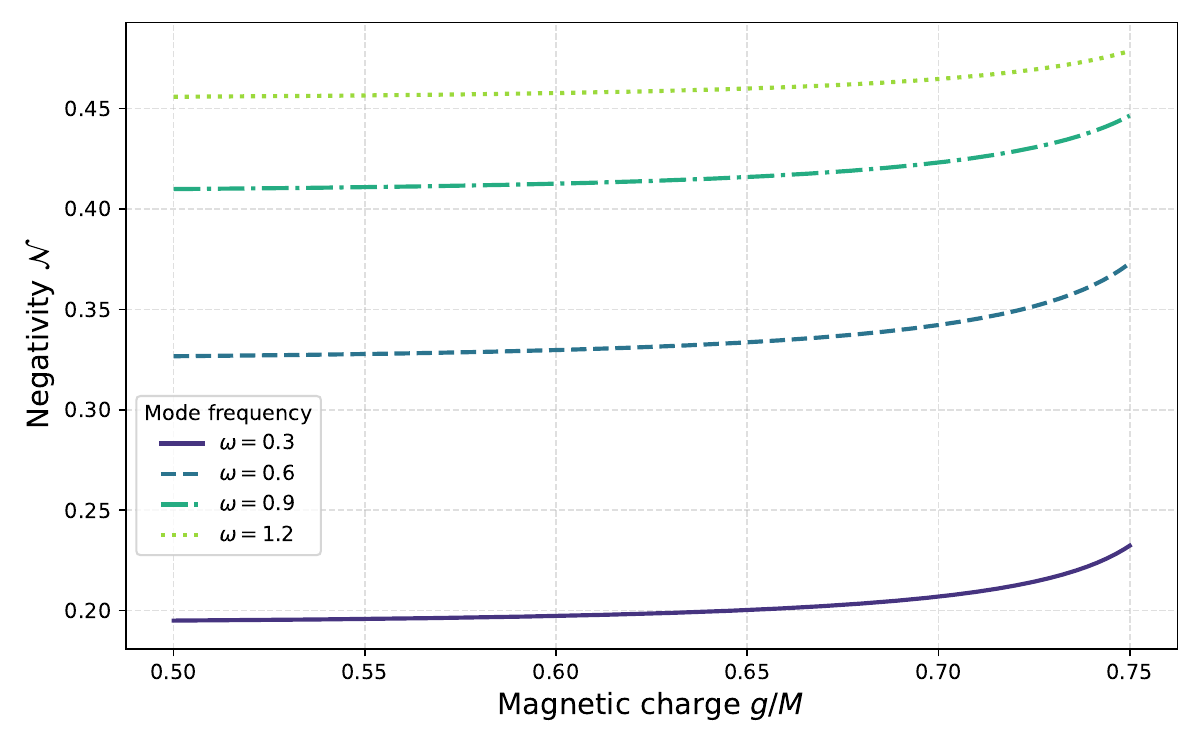}
    \caption{Negativity as a function of the magnetic charge $g$ for selected frequencies $\omega$.}
    \label{fig:nega_bardeen_g}
\end{figure}

We now explore the fate of the Alice-Rob bipartite state in the regular Bardeen black hole, whose only additional parameter is the magnetic charge $g$. Figure~\ref{fig:nega_bardeen_g} collects the negativity curves for several mode frequencies $\omega$ at fixed $\rho$.

Increasing $g$ weakens the near-horizon thermal bath and therefore reduces the degradation of distillable entanglement. 

Unlike the RN case, which exhibits a shallow minimum in the negativity, the Bardeen curves increase monotonically. Near the upper bound on the magnetic charge, $g/M\approx0.75$, the outer and inner horizons approach one another, the negativity reaches its maximum, and entanglement is degraded the least.

As a consequence, in contrast with the RN case, the entanglement is always less degraded than in the Schwarzschild configuration.

\subsection{Results for black holes with de~Sitter phases}\label{sec:results2}

\begin{figure}[H]
    \centering
    \includegraphics[width=1\linewidth]{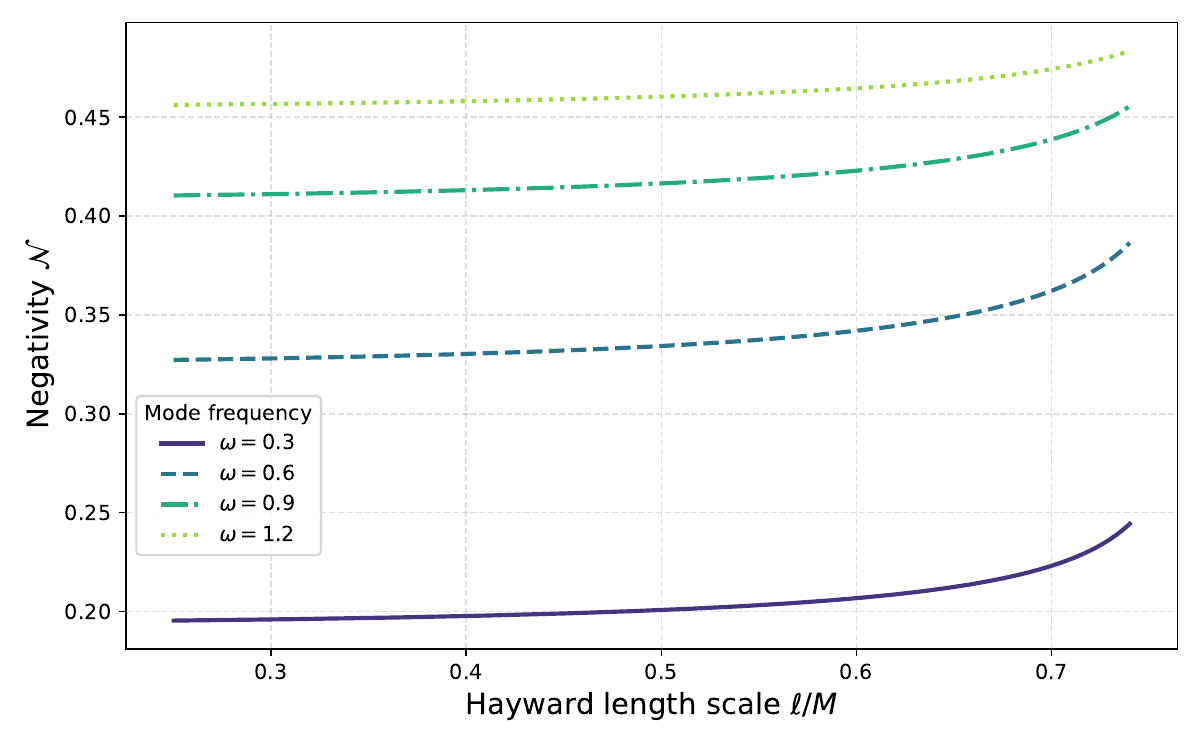}
    \caption{Negativity as a function of the length parameter $\ell$, for various mode frequencies $\omega$.}
    \label{fig:nega_hayward_ell}
\end{figure}

Figure~\ref{fig:nega_hayward_ell} shows the negativity as a function of the regularizing length-scale parameter $\ell/M$. These results are very similar to the Bardeen ones. Defining the Bardeen and Hayward negativity functions respectively as $\mathcal{N}^{\mathrm{B}}(x)$ and $\mathcal{N}^{\mathrm{H}}(x)$, we empirically observe that
\begin{equation}
    \mathcal{N}^{\mathrm{B}}(x)\approx \mathcal{N}^{\mathrm{H}}(x-0.25).
\end{equation}

Including the one-loop correction approximately shifts every curve downward while preserving its shape. This behavior is consistent with a slight increase in the effective temperature and therefore with enhanced entanglement degradation.

In cases like this one, that is, when Rob's proper acceleration is monotonic, there exists a one-to-one relation between Rob's proper acceleration and the Hawking temperature. Therefore, the entanglement loss can be traced back to the temperature.

\begin{figure}[H]
  \centering
  \includegraphics[width=1\linewidth]{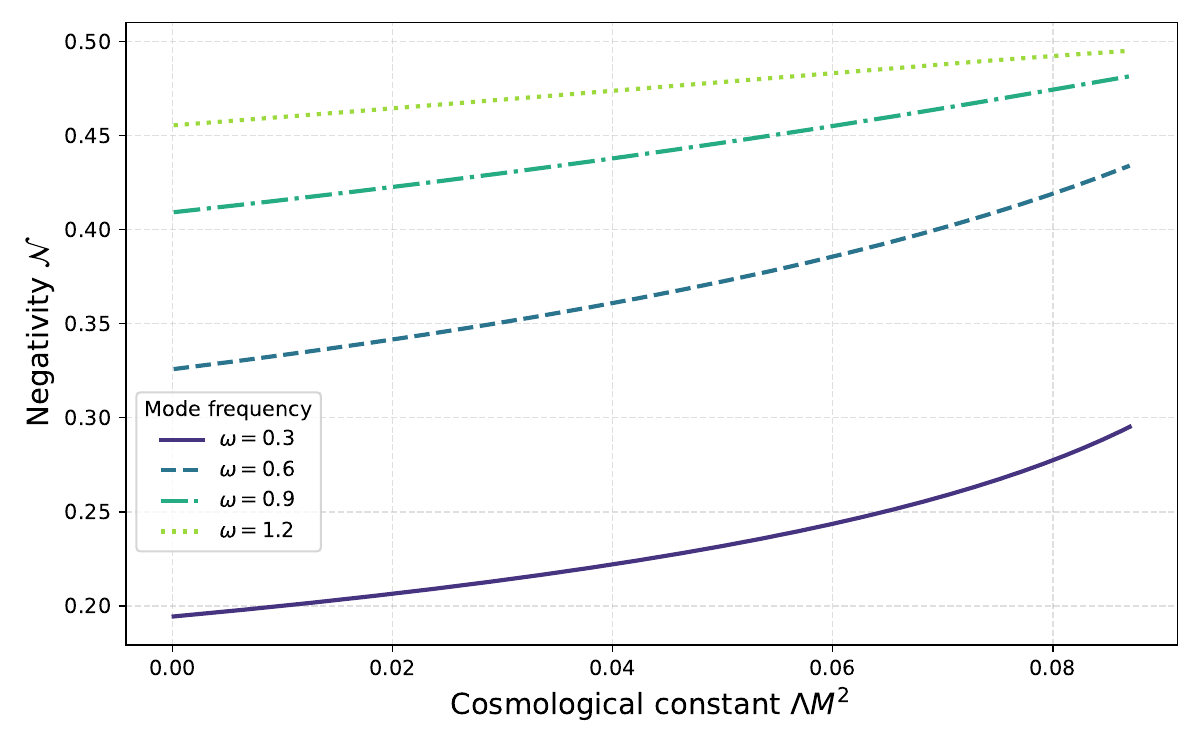}
  \caption{Negativity as a function of the dimensionless combination $\Lambda M^{2}$ at fixed $\rho$. Increasing $\Lambda$ lowers the surface gravities $\kappa_\pm$ of the black hole and cosmological horizons, and hence their Hawking temperatures, so thermal noise is suppressed and the negativity rises as the bound $9\Lambda M^{2}\approx 0.087$ is approached.}
  \label{fig:nega_sds_lambda}
\end{figure}

\subsection{Results for the Schwarzschild-de Sitter spacetime}

As mentioned in Sec.~\ref{sec:de_sitt_bh}, the relevant horizon in this case is the inner one. The physically relevant region of space is the cavity $r_- < r < r_+$. When $\Lambda$ is small, this region is large, but it shrinks as $\Lambda$ increases. The upper bound, computed with Eq.(\ref{eq:rindler_validity_condition}), turns out to be $9\Lambda M^2 = 0.087$. This ensures both that the Rindler approximation around the inner horizon remains valid and that Rob stays inside the cavity, far enough from the outer cosmological horizon. For these reasons, and given that the inner horizon is not a Cauchy horizon, it is natural to base the discussion of entanglement degradation on the thermodynamics of the inner horizon.

We now analyze the SdS case. Results are shown in Fig.\ref{fig:nega_sds_lambda}. In the Schwarzschild limit, $\Lambda \to 0$, the Hawking temperature associated with the relevant inner horizon is maximal within this family, so Rob is immersed in the strongest thermal bath considered in the SdS scan. The resulting thermal mixing strongly degrades the original bipartite entanglement with Alice. As $\Lambda$ increases, the relevant horizon temperature decreases monotonically and the negativity rises smoothly.

Compared with the other spacetimes considered, the SdS case is the one in which high-frequency modes come closest to the inertial negativity value $1/2$ near the upper bound. This trend should be interpreted cautiously, however, because the Rindler approximation is least predictive in that regime. Nevertheless, in the other considered spacetimes, the corresponding near-bound values remain farther from $1/2$.

For any fixed $\Lambda$, the lowest-frequency modes are most affected by thermal excitations and are therefore the least entangled. By contrast, higher-frequency modes are more robust and reach the plateau more rapidly. This is the same hierarchy observed near Schwarzschild horizons and in all other cases considered.

\subsection{Impact of regular black hole parameter on entanglement degradation}\label{sec:impact}

We now summarize the results from the previous subsections to clarify how each parameter controls the loss of entanglement between the inertial and static observers. The quantity of interest is the negativity $\mathcal N$ introduced in Eq.~\eqref{eq:bosonic_negativity}, evaluated at fixed $\rho$ and dimensionless mode frequency $\omega$.

In the RN spacetime, the interplay between charge and mass produces a shallow minimum in $\mathcal N$ as $Q/M$ is increased from $0$; see Fig.~\ref{fig:nega_rn_charge}. The location of the dip is determined entirely by $\rho$ via Eq.~\eqref{eq:q/m_min}. For $\tfrac12(1+\sqrt{5-2\sqrt3})\le\rho\le\tfrac12(1+\sqrt{5+2\sqrt3})$, the radicand in Eq.~\eqref{eq:q/m_min} is negative, the proper acceleration has no local maximum, and hence the negativity has no minimum. We also find that the RN case is the only one in which the negativity drops below the Schwarzschild value. This indicates that sufficiently small charges can further degrade entanglement, whereas sufficiently large charges preserve it much better than in Schwarzschild. 

In all other geometries, the proper acceleration is monotonic and the negativity remains above the Schwarzschild baseline. 

In both the Bardeen and Hayward geometries, the negativity varies monotonically with the corresponding parameter; see Figs.~\ref{fig:nega_bardeen_g} and \ref{fig:nega_hayward_ell}. As the upper bound is approached, $\mathcal{N}$ saturates at a value determined by the finite acceleration. We also find an empirical relation between the two geometries: $\mathcal{N}^{\mathrm{B}}(x)\approx \mathcal{N}^{\mathrm{H}}(x-0.25)$. For the Hayward case, the one-loop correction acts predominantly as an approximately uniform downward shift of the curves while leaving their shapes essentially unchanged. This behavior is consistent with additional gravitational effects from vacuum polarization further enhancing the Hawking temperature.

In the SdS case, the positive cosmological constant also cools the relevant black hole horizon. Although the results are qualitatively similar to those obtained for the other geometries, there is a key difference: Rob is located between the black hole and cosmological horizons, whereas in all other cases he is outside the outer horizon.

For these reasons, making a direct comparison between the SdS and Schwarzschild results is not possible.

Nevertheless, as discussed, the cavity between the horizons is the spacetime region relevant for the Hawking temperature and the associated entanglement degradation. We observe that increasing $\Lambda M^2$ lowers the surface gravity, thereby suppressing thermal noise and yielding, among all the geometries considered, the greatest recovery of entanglement for high-frequency modes (Fig.~\ref{fig:nega_sds_lambda}).

Taken together, these results suggest that entanglement negativity could, in principle, help distinguish the standard Schwarzschild solution from alternative regular or non-vacuum black hole geometries, and in between those regular or non-vacuum black hole geometries themselves. Assessing observational feasibility, however, is outside the scope of the present analysis.

\section{Final outlooks}\label{sec:discussion}

In this work, we investigated how the characteristic parameters of regular black hole spacetimes, from the magnetic charge to the scale of the de Sitter core, affect entanglement degradation in a bipartite scalar field system. We also compared these results with the corresponding standard black hole solutions, specifically the RN and SdS geometries, which respectively incorporate electric charge and a de Sitter phase.

Throughout, we employed a \emph{local Rindler approximation} for an observer located close to the event horizon, denoted Rob. We then related the near-horizon geometry of each spacetime to the entanglement loss experienced by a state shared by Rob and an inertial observer, Alice.

We extended the method to generic spherically symmetric spacetimes. In particular, we highlighted the role of $r_0$, the distance from the horizon, as the parameter controlling the range over which the local Rindler description can be trusted. Then we computed the validity bound of the local Rindler approximation in near extremal configurations.

Because these regular black holes admit a Hawking temperature, our analysis shows that entanglement degradation is not determined solely by the presence of horizons. Rather, it is also modulated by the free parameters of the underlying geometry, a feature that emerges clearly from the comparative analysis.

More specifically, we studied the Bardeen, Hayward, and generalized Hayward spacetimes as regular black hole models. In the Bardeen case, the parameter $g$ acts as a magnetic charge that regularizes the solution at $r=0$. In the Hayward class, the parameter $\ell$ sets the scale of an effective de Sitter core, which likewise removes the central singularity.

These solutions were compared with the Schwarzschild spacetime as the simplest reference case. In addition, RN and SdS were considered as explicit counterparts to the Bardeen and Hayward families, respectively.

In the RN case, we identified a non-monotonic behavior of the negativity. For certain values of $\rho$, the geometry yields a local maximum in Rob's proper acceleration, which is mirrored by a local minimum in the entanglement. Within our setup, the charge value corresponding to maximum acceleration therefore corresponds to the least favorable RN configurations for quantum communication.

For the regular black holes considered here, namely the Bardeen and Hayward models, the entanglement degradation is monotonic. The negativity curves in the Bardeen and Hayward geometries are very similar; empirically, an approximate translation in parameter space brings them into near coincidence, with discrepancies of only a few percent.

Among the cases analyzed, SdS is the only configuration for which an almost complete recovery of the entanglement appears possible at high frequencies. In this case, since Rob is sitting in the cavity between the inner and outer horizon, a direct comparison with the Schwarzschild result is not possible.

Several directions remain open for future work. Our analysis relied on the single Unruh mode approximation, but could be extended to localized wave packets to provide a more realistic description \cite{Bruschi2010}. Another natural extension is to study multipartite entanglement, for example, in GHZ or W states, in the same backgrounds. This may clarify whether higher-order correlations are more or less robust than the bipartite entanglement considered here. More broadly, entanglement degradation may serve as an additional diagnostic for distinguishing mimickers from black holes, especially in combination with future observational constraints. Thus, exploring further regular metrics \cite{Bambi:2023try} in combination with external fields \cite{Luongo:2018lgy} and additional exotic fields \cite{DAgostino:2021vvv} will shed light on how near-horizon geometry is modified accordingly.

\section*{Acknowledgments}
OL acknowledges  Hernando Quevedo for discussions on the topic of Rindler space applied to the contexts of entanglement degradation. He is also grateful to Daniele Malafarina and Roberto Giambò for stimulating debates on alternatives to de Sitter space. SM acknowledges financial support from COST Action CA23115: Relativistic Quantum Information, funded by COST (European Cooperation in Science and Technology) and from PNRR Italian
Ministry of University and Research project PE0000023-NQSTI.

\bibliographystyle{abbrv}
\bibliography{bibliography}

\end{document}